\documentclass[iop,apj,numberedappendix,onecolumn]{emulateapj}
\usepackage{natbib}
\usepackage{graphicx}
\usepackage{amsmath,amsfonts}
\usepackage{epstopdf}

\slugcomment{Accepted for publication in ApJ, April 1, 2013}

\defcitealias{IPDF}{JG12}

\def\lsim{\mathrel{\raise.3ex\hbox{$<$\kern-.75em\lower1ex\hbox{$\sim$}}}}
\def\gsim{\mathrel{\raise.3ex\hbox{$>$\kern-.75em\lower1ex\hbox{$\sim$}}}}
\def\propsim{\mathrel{\raise.3ex\hbox{$\propto$\kern-.75em\lower1ex\hbox{$\sim$}}}}

\begin{document}
\title{ Interferometric Visibility of a Scintillating Source: Statistics at the Nyquist Limit }
\shorttitle{Visibility Scintillation Statistics}

\author{M. D. Johnson and C. R. Gwinn}
\shortauthors{Johnson \& Gwinn}
\affil{Department of Physics, University of California, Santa Barbara, CA 93106, USA}
\email{ michaeltdh@physics.ucsb.edu,cgwinn@physics.ucsb.edu}

\begin{abstract}
We derive the distribution of interferometric visibility for a source exhibiting strong diffractive scintillation, with particular attention to spectral resolution at or near the Nyquist limit. We also account for arbitrary temporal averaging, intrinsic variability within the averaging time, and the possibility of spatially-extended source emission. We demonstrate that the interplay between scintillation and self-noise induces several remarkable features, such as a broad ``skirt'' in the visibility distribution. Our results facilitate the interpretation of interferometric observations of pulsars at meter and decimeter wavelengths.
\end{abstract}


\section{Introduction}

Radio astronomy is now exploring two limiting regimes. 
First, with the construction of ever-larger collecting areas, noise intrinsic to the source (i.e.\ \emph{self-noise}) can be a major, or even dominant, component of the noise budget.
Second, observing systems at decimeter wavelengths can now record baseband data over wide bandwidths and with sufficient bit-rates to effectively capture the full information content of the signal (up to the fundamental limit described by the Nyquist sampling theorem). These advances motivate mathematical descriptions of the new regimes and the development of novel techniques that optimally utilize the signal information.

These considerations prompted a previous work \citep[hereafter JG12]{IPDF}, in which we derived the probability density function (PDF) of flux density, sampled at the Nyquist rate, for a scintillating source. We accounted for background noise\footnote{By \emph{background noise}, we mean the sum of all noise other than that of the source (e.g., receiver, spillover, atmosphere, and background sky noise).}, 
arbitrary temporal averaging, the possibility of decorrelation of the scintillation pattern within the averaging time, and spatially-extended source emission. We also outlined tests to identify self-noise, as well as a mechanism to detect rapid intrinsic variability of a signal. We then demonstrated the success of this description on 800 MHz observations of the Vela pulsar and, thereby, obtained a spatial resolution of approximately 4 km at the pulsar \citep{GUPPI_Size_2}.

We now extend these mathematical results and techniques to interferometric visibility, which preserves much of the statistical framework. 
This extension involves a pair of stations, which observe independent background noise but correlated scattering kernels and identical source noise. However, these mild modifications lead to substantial changes in the analytical results for the distribution functions. Furthermore, the interferometric visibility is complex, and thus is characterized by a two-dimensional PDF, whereas the flux density is a positive real number. 
We frequently utilize projections as a natural mechanism for exploring the transition between these domains. 
We also offer several other close analogs with common metrics for flux-density, and we provide tools to quantify the complex noise in sampled visibilities.

Pulsars are some of the richest targets for interferometry, and we have tailored our treatment to account for their extraordinary variability. In particular, we emphasize results for single-pulse studies with Nyquist-limited resolution. We then connect to traditional limits through asymptotic forms of our results, as the averaging is increased and self-noise becomes negligible. However, our principal goal is to facilitate precision tests using interferometric observations, particularly of pulsars, that can robustly distinguish between intrinsic and extrinsic characteristics and can sensitively probe delicate modifications of either. 

\subsection{Assumptions and Strategy for Comparison with Observations}
\label{sec::obs_strategy}

Our scope is deliberately broad, intended to encompass the majority of pulsar observations at meter and decimeter wavelengths. However, to assess the suitability for comparison with any particular observation, we now outline the assumptions underlying our results. Our \emph{physical} requirements are that
\begin{itemize}
\item The source emits amplitude-modulated noise \citep{Rickett_AMN}.
\item The scattering is strong. That is, the diffractive scale $r_{\rm d} = \lambda/\theta$ is much smaller than the Fresnel scale $r_{\mathrm F} = \sqrt{\lambda D}$, where $\lambda$ is the observing wavelength, $\theta$ is the angular size of the scattering disk, and $D$ is the characteristic distance to the scattering material \citep{Cohen_Cronyn}.
\end{itemize}
The first assumption is quite robust for most, if not all, astrophysical sources, simply because the superposition of many independent radiators will produce a signal of this form. Also, the majority of meter and decimeter pulsar observations fall easily within the regime of strong scattering: $r_{\rm d}/r_{\rm F} \ll 1$.

On the other hand, the \emph{instrumental} assumptions that we require are that
\begin{itemize}
\item The data explore a large representation of the full ensemble of diffractive scintillation: 
\mbox{$(B /\Delta \nu_{\mathrm{d}} ){\times}(t_{\mathrm{obs}}/\Delta t_{\mathrm{d}}) \gg 1$}, where $\{ B,\, t_{\rm obs} \}$ are the total observational bandwidth and duration, and $\{\Delta \nu_{\rm d},\,\Delta t_{\rm d}\}$ are the characteristic bandwidth and timescale of the scintillation pattern.
If $B /\Delta \nu_{\mathrm{d}} \gg 1$, then intrinsic single-pulse amplitudes may be estimated, allowing a comparison between the data and our models that requires no fitted parameters.
\item The data are coherently dedispersed, baseband shifted, and ``fringed'' with an appropriate phase model.
\item The time to form each spectrum, or the \textit{accumulation time} $t_{\rm acc}$, is much longer than the pulse-broadening timescale $t_0$ but much shorter than the characteristic scintillation timescale $\Delta t_{\rm d}$.
\end{itemize}

The last assumption is perhaps the only one that is atypical of modern observations and processing. However, it ensures that the spectra reflect the convolution action of the scattering on the intrinsic pulsar signal with a stochastic ``propagation kernel'' \citep{Hankins_Coherent,Williamson,Intermittent_Noiselike_Emission}. For pulsars, a particularly elegant limiting case is the formation of spectra that include \emph{all} the pulsed power. 
This convolution leads to extremely general results; we do not assume thin-screen scattering or a particular spectrum for the density inhomogeneities of the interstellar medium (ISM), for example.

Our work also addresses the physical possibility of an extended emission region and the instrumental possibility of averaging (or integration) of the calculated visibilities. For the former, we assume that the source is a small fraction of the magnified diffractive scale (see \S\ref{sec::PDF_Emission_Size}); for the latter, we assume that the averaging is of $N$ independent cross-spectra (say, from different pulses), over a timescale much shorter than $\Delta t_{\rm d}$ (i.e.\ the ``snapshot image'' of \citet{NarayanGoodman89_1}). While frequency averaging is analogous, it inherits additional information from non-stationary signals and requires the specification of individual pulse profiles, whereas a description of temporal averaging requires only the phase-averaged source intensity for each pulse (see \citetalias{IPDF} for details). Hereafter, we exclusively use $N$ to denote this degree of temporal averaging.

\subsection{Relation to Previous Work}

For non-scintillating sources, the statistics of interferometric visibility are well-known; see \citet{Moran_76} or \citet{TMS}. In particular, the self-noise has been carefully characterized. For example, \citet{Kulkarni_SelfNoise} analyzed the noise in synthesis imaging for sources of arbitrary strength, while \citet{Anantharamaiah} studied the noise, with an emphasis on extremely bright sources. 

Scintillation complicates the statistics, and even the noise-free (i.e.\ infinite-averaging) case warrants careful attention. \citet{NarayanGoodman89_1} and \citet{GoodmanNarayan89_2} analyzed this limit via both numerical and analytical techniques, with an emphasis on moments of the distribution of visibility. Also, \citet{IVSS} calculated the full PDF of interferometric visibility for a scintillating source in this limit and accounted for the effects of an extended emission region. 

\citet{Vela_noise} and \citet{Vela_18cm} then incorporated the contribution of both self-noise and background noise for modest averaging ($N \gsim 20$) by applying the central limit theorem to the averaged visibilities. The present work extends these ideas by adopting a different strategy, following \citetalias{IPDF}. Namely, we establish the statistics for individual spectral samples and then determine the effects of averaging by a convolution of independently drawn samples. This method permits the treatment of data with arbitrary averaging, $N$, including no averaging (i.e.\ $N=1$), and requires no assumptions about the nature of the intrinsic variability or of the scattering material. 

To put our results in the context of these earlier efforts, we also present asymptotic forms of our equations for strong signals and high degrees of averaging. However, our principal goal is to facilitate direct comparisons with observations.


\subsection{Outline of Paper}
In \S\ref{sec::VPDF}, we focus on visibility statistics for a scintillating point source. We account for temporal averaging, but assume that the scintillation pattern is fixed within each average. Under these constraints, we first outline the essential statistical framework for the visibility statistics (\S \ref{sec::Field_Statistics}) and calculate moments and noise of the visibility statistics within a fixed scintillation element (\S\ref{sec::Snapshot_Moments}). We then outline approximation schemes for the PDF of visibility during a fixed scintillation element (\S\ref{sec::Snapshot_Approx}) and demonstrate how to calculate the PDF of visibility after including the scintillation ensemble (\S\ref{sec::V_PDF}). We give examples of the visibility statistics in limiting regimes (\S\ref{sec::Examples}) and demonstrate that the combination of self-noise and scintillation introduces a ``skirt'' in the visibility PDF that dominates its asymptotic form, regardless of the source strength or the baseline (\S\ref{sec::Asymptotic}). We also provide a prescription for estimating the self-noise (\S\ref{sec::VSelfNoise}).

Next, in \S\ref{sec::PDF_Emission_Size}, we quantify the influence of a spatially-extended emission region on this PDF and discuss the potential of interferometry to resolve the size and anisotropy of such emission. We demonstrate that the effect of a small emission region depends only on two parameters, regardless of the scattering geometry; however, the translation to a dimensionful size at the source depends on the scattering geometry. We also derive a simplified version of the PDF of visibility for the special case of a zero-baseline interferometer.

Finally, in \S\ref{sec::Summary_V}, we summarize our results and outline some observational prospects.

\section{PDF of Visibility}
\label{sec::VPDF}

We now derive the expected PDF of interferometric visibility arising from a scintillating point source. We allow arbitrary temporal averaging, in sets of $N$ independent cross-spectra, but assume that the scintillation pattern is fixed during each average. All of our results apply to scalar electric fields (i.e.\ a single linear or circular polarization) that conform to the assumptions described in \S\ref{sec::obs_strategy}.

Our notation follows \citetalias{IPDF}. Namely, $z_x$ denotes a circular complex Gaussian random variable with unit variance, indexed by $x$; $G_x$ denotes an exponential random variable with unit scale. We use $P()$ to generically denote a PDF with respect to the given variables and parameters, and we present PDFs $P(w)$ of complex quantities $w$ with respect to the metric $d\mathrm{Re}[w] d\mathrm{Im}[w]$. Occasionally, we employ the shorthand $w_{\rm r} \equiv \mathrm{Re}(w)$, $w_{\rm i} \equiv \mathrm{Im}(w)$.

Furthermore, to help visualize the visibility PDF, we rely on projections. These projections are most enlightening along the real axis $V_{\rm r} \equiv \mathrm{Re}(V)$, where they emphasize the relative effects of differing baselines. We favor the following two projections:
\begin{align}
P(V_{\rm r};N) &\equiv \int dV_{\rm i}\,  P(V;N),\\
\nonumber Q(V_{\rm r};N) &\equiv \int dV_{\rm i}\,  V_{\rm i}^2 P(V;N).
\end{align}
The first projection identifies the concentration of density toward greater real part, which reflects the average visibility; the second (weighted) projection quantifies the imaginary spread of density, which reflects the influence of scintillation and the noise from both the background and the source.

\subsection{Field Statistics}
\label{sec::Field_Statistics}

In \S3.1 of \citetalias{IPDF}, we derived the electric-field statistics for a scintillating point source. These statistics depend on the amplitude-modulated noise of the source, the strong scattering of the ISM, and the receiver noise. We now review the basic ingredients of this description and establish the necessary notation.

For the electric field, the amplitude-modulated noise takes the form $\sqrt{A_j I_{\mathrm{s}}} f_i \epsilon_i$. Here, $A_j$ is a dimensionless amplitude factor, indexed by pulse, that accounts for pulse-to-pulse variations, $I_{\mathrm{s}}$ is a constant characteristic scale of source intensity, $f_i$ is a power-preserving envelope, and $\epsilon_i$ is white Gaussian noise of unit variance. Thus, the pulse profile is simply $|f_i|^2\!$, the gated signal has mean amplitude $A_j I_{\rm s}$, and $\epsilon_i$ accounts for the noiselike nature of the emission. Note that this treatment accommodates arbitrary variability of the pulsar, such as the possibility of correlated pulse-to-pulse variations, log-normal amplitude statistics, or nanosecond-scale bursts \citep{Rickett_microstructure,Vela_lognormal,Kramer_Vela}.

During a fixed scintillation element, the scattering acts to convolve this intrinsic emission with a stochastic propagation kernel: $g_i \eta_i$. In the strong-scattering limit, the form of this kernel is similar to the emission of the pulsar. Namely, a power-preserving envelope $g_i$ modulates Gaussian noise $\eta_i$. This envelope is more commonly described by its squared norm: the pulse-broadening function $|g_i|^2$\!.

Finally, an observer samples the propagated signal in the presence of white background noise: $\sqrt{I_{\rm n}} \beta_i$. Here, $I_{\rm n}$ is a constant, characteristic scale of the background noise, and $\beta_i$ is white Gaussian noise of unit variance. If the background noise changes significantly with time, then a changing scale may be added, analogous to $A_j$.

The observed scalar electric-field time series $x_i$ and its Fourier-conjugate spectrum $\tilde{x}_i$ are thus given by
\begin{align}
\label{eq::E_I}
x_i = \sqrt{A_j I_{\mathrm{s}}} \left[ \left( f \epsilon \right) \ast \left(g \eta \right) \right]_i + \sqrt{I_{\mathrm{n}}}\beta_i \Rightarrow \tilde{x}_i &= \sqrt{A_j I_{\mathrm{s}}} \left( \tilde{f} \ast \tilde{\epsilon} \right)_i \left(\tilde{g} \ast \tilde{\eta} \right)_i + \sqrt{I_{\mathrm{n}}} \tilde{\beta}_i,
\end{align}
where a tilde denotes a Fourier conjugate variable.

Because $\tilde{\epsilon}_i$, $\tilde{\eta}_i$, and $\tilde{\beta}_i$ are mutually independent (circular complex Gaussian) white noise, a single spectral sample $\tilde{x}_i$ is of the form $\sqrt{A_j I_{\mathrm{s}}} z_{\mathrm{f}} z_{\mathrm{g}} + \sqrt{I_{\mathrm{n}}} z_{\mathrm{b}}$, where $z_{\mathrm{f}} \equiv (\tilde{f}\ast \tilde{\epsilon})_i$, $z_{\mathrm{g}}\equiv (\tilde{g} \ast \tilde{\eta})_i$, and $z_{\mathrm{b}} \equiv \tilde{\beta}_i$ are each circular complex Gaussian random variables with unit variance. If the scintillation is held fixed (i.e.\ $z_{\mathrm{g}} = \mathrm{const.}$), then the intensity $|\tilde{x}_i|^2$ is drawn from an exponential distribution with scale $\bar{I}_j \equiv A_j I_{\mathrm{s}} |z_{\mathrm{g}}|^2 + I_{\mathrm{n}}$. 

For interferometric visibility, the observer measures the covariance of the electric fields at two stations, which we denote by unprimed and primed variables: $V_i \equiv \tilde{x}_i \tilde{x}_i'^\ast$ \citep{TMS}. The electric fields at the two stations arise from identical intrinsic emission, $\tilde{f}' \ast \tilde{\epsilon}' = \tilde{f} \ast \tilde{\epsilon}$; however, a difference in sensitivity or gain between the stations will affect the overall scale: $I_{\mathrm{s}} \neq I_{\mathrm{s}}'$. The background noise is independent ($\langle z_{\rm b} z_{\rm b}'^\ast \rangle = 0$) and with different variance ($I_{\mathrm{n}} \neq I_{\mathrm{n}}'$) at the two stations. Finally, the stochastic part of the propagation will also differ, with a baseline-dependent correlation $\rho_{\rm g} \equiv \left \langle z_{\rm g} z_{\rm g}'^\ast \right \rangle$.

Hence, the average of $N$ visibilities from different pulses takes the form
\begin{align}
\label{eq::V_rV}
V = \frac{1}{N} \sum_{j=1}^N &\left(\sqrt{A_j I_{\mathrm{s}}} z_{\mathrm{f},j} z_{\rm g} + \sqrt{I_{\mathrm{n}}} z_{\mathrm{b},j} \right)
\left(\sqrt{A_j I_{\mathrm{s}}'} z_{\mathrm{f},j} z_{\rm g}' + \sqrt{I_{\mathrm{n}}'} z_{\mathrm{b},j}' \right)^\ast\!.
\end{align}
Even in the zero-baseline limit ($z_{\rm g}' = z_{\rm g}$), Eq.\ \ref{eq::V_rV} differs from the corresponding intensity result because of the assumption of independent background noise at the stations.

As Eq.\ \ref{eq::V_rV} shows, the measured visibility is the average of $N$ random variables, each of which is the product of two correlated circular complex Gaussian random variables with respective variances $\bar{I}_j \equiv A_j I_{\mathrm{s}} |z_{\rm g}|^2 + I_{\mathrm{n}}$ and $\bar{I}_j' \equiv A_j I_{\mathrm{s}}' |z_{\rm g}'|^2 + I_{\mathrm{n}}'$ and correlation 
\begin{align}
\rho \equiv  z_{\rm g} z_{\rm g}'^\ast A_j \sqrt{\frac{I_{\mathrm{s}} I_{\mathrm{s}}'}{\bar{I}_j\bar{I}_j'}}.
\end{align}
This correlation changes with scintillation and is complex because of covariance between the real and imaginary part of electric fields at the two stations. 

For pulsar observations, one can examine the mean of the off-pulse and on-pulse spectra at each station to estimate the parameters $I_{\rm s}$, $I_{\rm s}'$, $I_{\rm n}$, $I_{\rm n}'$, and $A_j$. These measurements fully characterize the visibility statistics of Eq.\ \ref{eq::V_rV}.

\subsection{Moments and Noise of Snapshot Visibilities}
\label{sec::Snapshot_Moments}
Using Eq.\ \ref{eq::V_rV}, we can evaluate moments of the visibility distribution for a ``snapshot image'' (i.e.\ the scintillation variables $z_{\rm g}$ and $z_{\rm g}'$ are held fixed). These moments include the effects of pulsar variability and self-noise. For example, 
\begin{align}
\label{eq::V_moments}
\langle V \rangle &= \langle A \rangle_N \sqrt{I_{\rm s} I_{\rm s}'} z_{\rm g} z_{\rm g}'^\ast \\
\nonumber \left \langle \mathrm{Re}(V)^2 \right \rangle &= 
\frac{I_{\rm n} I_{\rm n}'}{2N}
+ \frac{\langle A \rangle_N}{2N} \left( I_{\rm s} I_{\rm n}' |z_{\rm g}|^2 + I_{\rm s}' I_{\rm n} |z_{\rm g}'|^2  \right)
+ \left( \frac{\langle A^2 \rangle_N}{N} + \langle A \rangle_N^2 \right) I_{\rm s} I_{\rm s}' \mathrm{Re}\left[ z_{\rm g} z_{\rm g}'^\ast \right]^2\\
\nonumber \left \langle \mathrm{Im}(V)^2 \right \rangle &= 
\frac{I_{\rm n} I_{\rm n}'}{2N}
+ \frac{\langle A \rangle_N}{2N} \left( I_{\rm s} I_{\rm n}' |z_{\rm g}|^2 + I_{\rm s}' I_{\rm n} |z_{\rm g}'|^2  \right)
+ \left( \frac{\langle A^2 \rangle_N}{N} + \langle A \rangle_N^2 \right) I_{\rm s} I_{\rm s}' \mathrm{Im}\left[ z_{\rm g} z_{\rm g}'^\ast \right]^2.
\end{align}
These expressions present ensemble averages over the noise of the pulsar and background, while the set of $N$ pulse amplitudes $\{ A_j \}$ and the scintillation factors are held fixed; $\langle \hdots \rangle$ denotes the average over noise, whereas $\langle \hdots \rangle_N$ denotes the average over the $N$ pulse amplitudes.

We can also quantify the noise using these moments. 
Within a single scintillation element, the noise takes the form
\begin{align}
\label{eq::V_noise}
\left \langle |\delta V|^2 \right \rangle &\equiv \left \langle \left| V - \langle V \rangle \right|^2 \right \rangle\\
\nonumber &=
\frac{I_{\mathrm{n}} I_{\mathrm{n}}'}{N}
+ \frac{1}{N} \left( \left|\frac{z_{\rm g}}{z_{\rm g}'}\right| \sqrt{\frac{I_{\mathrm{s}}}{I_{\mathrm{s}}'}} I_{\mathrm{n}}' + \left|\frac{z_{\rm g}'}{z_{\rm g}}\right| \sqrt{\frac{I_{\mathrm{s}}'}{I_{\mathrm{s}}}} I_{\mathrm{n}}\right) \left| \langle V \rangle \right|
+ \frac{1}{N} \left(1 + \frac{\left \langle \delta A^2 \right \rangle_N}{\langle A \rangle_N^2} \right) \left| \langle V \rangle \right|^2 \\
\nonumber &\equiv 2b_0 + 2 b_1 \left| \left \langle V \right \rangle \right| + b_2 \left| \left \langle V \right \rangle \right|^2\!.
\end{align}
The noise is a quadratic function of the signal; hence, pulsar field statistics are heteroscedastic \citep{Noise_Inventory,Vela_noise}. In particular, this equation demonstrates the contribution of the source to the noise, as originally investigated by \citet{Dicke} and described by the familiar radiometer equation. Observe that, for the special case of a zero-baseline interferometer, the noise coefficients, $b_i$, are independent of the particular scintillation element $\{ z_{\rm g},z_{\rm g}'\}$.

\begin{figure}[t]
\centering
\includegraphics*[width=1.0\textwidth]{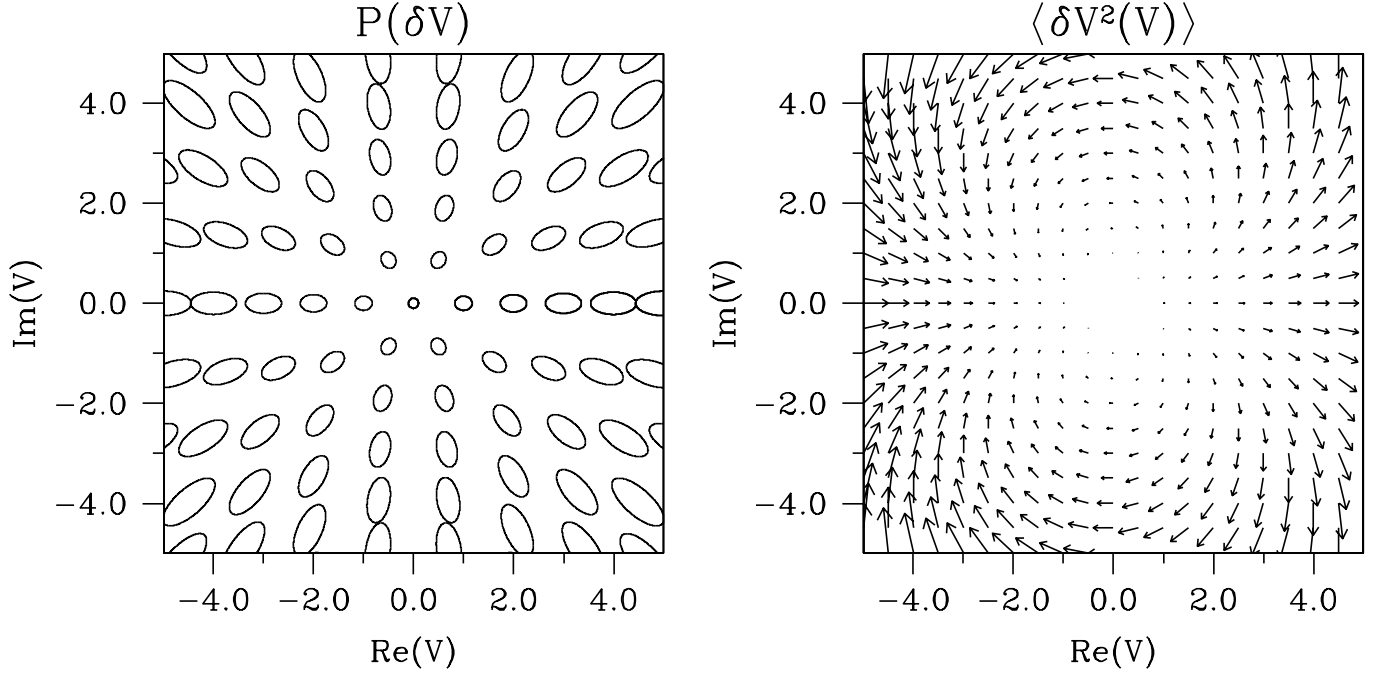}
\caption{
Characteristics of complex noise $\delta V$ for snapshot visibilities. (\textit{left}) Ellipses denote the standard deviation of the Gaussian distribution of noise, centered on the mean visibility. The major axis is oriented with the phase of the visibility and scales quadratically with the signal, while the minor axis scales linearly with the signal (see \S\ref{sec::Snapshot_Moments}). The plotted noise corresponds to noise on a short baseline with all source and background intensities unity, and $N=100$. (\textit{right}) Average complex noise $\langle\delta V^2(V)\rangle$, estimated as described in \S\ref{sec::VSelfNoise}. The increasing vector length with $|V|$ shows the increasing influence of self-noise, while the change in direction reflects the varying orientation of the noise ellipse.
}
\label{fig::V_Noise}
\end{figure}

We can similarly obtain the variances parallel and perpendicular to the mean visibility:
\begin{align}
\label{eq::V_noise_pp}
\left \langle \delta V_{\parallel}^2 \right \rangle &= b_0 + b_1 \left| \left \langle V \right \rangle \right| + b_2 \left| \left \langle V \right \rangle \right|^2\!, \\
\nonumber \left \langle \delta V_{\perp}^2 \right \rangle &= b_0 + b_1 \left| \left \langle V \right \rangle \right|.
\end{align}
Thus, the noise scales quadratically in phase with the signal but only scales linearly at quadrature to the signal. Figure \ref{fig::V_Noise} illustrates this noise behavior.

Because they neatly separate the contributions of the background and source, the noise coefficients provide a valuable mechanism to study and quantify the various types of noise.
The self-noise coefficient, $b_2$, is particularly useful because it contains information about the intrinsic variability, which does not affect the average spectrum or correlation function \citep{Intermittent_Noiselike_Emission}. Other types of noise, such as quantization noise \citep{Cole_68,TMS,Jenet_Anderson}, will modify the coefficients but preserve the quadratic form \citep{Gwinn_Quant_Noise}. In \S\ref{sec::VSelfNoise}, we describe a procedure to estimate the self-noise.

\subsection{Approximating the Visibility Statistics within a Scintillation Snapshot}
\label{sec::Snapshot_Approx}
We now derive the PDF of visibility for samples that are collected and averaged within a single scintillation snapshot, characterized by $\{z_{\rm g},z_{\rm g}'\}$. 
Because the field statistics take a rather general form, $P(V|z_{\rm g},z_{\rm g}')$ is simply the $N$-fold convolution of the distribution of the product of correlated complex Gaussian random variables; we relegate the derivation and details of this distribution to the appendix (in particular, \S\ref{sec::Visibility_Average}).

We also present several approximation strategies, because the convolution of visibilities that are not statistically identical has no convenient analytical form. These strategies provide accuracy that is sufficient for most applications and constitute the analytical foundation for our subsequent results. Moreover, the first approximation that we derive, the i.i.d.\ approximation, is exact for $N=1$.

\subsubsection{The i.i.d.\ Approximation}
\label{sec:iid}

We derive our first approximation by assuming that the averaged visibilities for each set of $N$ pulses are independent and identically distributed (i.i.d.). To achieve this condition, we treat the pulse amplitudes as constant for each set of $N$ averaged visibilities: $A_j \mapsto A \equiv \langle A \rangle_N$. 
Nevertheless, this approximation preserves some information about the pulsar variability; the i.i.d.\ approximation is exact when $N=1$, for instance.

After this replacement, Eq.\ \ref{eq::V_Niid} gives the PDF of visibility:
\begin{align}
\label{eq::V_noscint}
P(V;N|z_{\rm g},z_{\rm g}') = \frac{N^{N+1}}{2^N \pi (N-1)!} \frac{\left( 1 - |\rho|^2 \right)^N }{a^{N+1}} |V|^{N-1}
\text{K}_{N-1}\left( N \frac{|V|}{a} \right) \exp \left( N \frac{\mathrm{Re}\left[V \rho^\ast \right]}{a} \right),\\
\nonumber 
\rho \equiv z_{\rm g} z_{\rm g}'^\ast A \sqrt{\frac{I_{\mathrm{s}} I_{\mathrm{s}}'}{\bar{I} \bar{I}'}},\qquad 
a \equiv \frac{\left(1-|\rho|^2\right)}{2} \sqrt{\bar{I} \bar{I}'},\qquad
\bar{I}  \equiv A I_{\mathrm{s}}  |z_{\rm g}|^2  + I_{\mathrm{n}} ,\qquad
\bar{I}' \equiv A I_{\mathrm{s}}' |z_{\rm g}'|^2 + I_{\mathrm{n}}'. 
\end{align}
In this expression, $K_N(x)$ is the modified Bessel function of the second kind. This approximation and its equivalent for intensity \citepalias[\S3.2.1]{IPDF} provide convenient tools for analytic work.

\subsubsection{The Gaussian Approximation}
\label{sec::Gaussian_Approximation_V}

As the number of averaged samples $N \rightarrow \infty$ within a fixed scintillation pattern, the PDF of visibility approaches an elliptical complex Gaussian distribution with mean $V = \langle A \rangle_N \sqrt{I_{\mathrm{s}}I_{\mathrm{s}}'} z_{\rm g} z_{\rm g}'^\ast$ and variances determined by Eq.\ \ref{eq::V_noise_pp}. \citet{Vela_noise} and \citet{Vela_18cm} described this limit and result, and verified the noise prescription using observations of the Vela pulsar.

In general, the linear term, $b_1$, of the noise polynomial depends weakly on the scintillation $\{ z_{\rm g},z_{\rm g}' \}$; however, for a short baseline, the noise ellipse depends only on the mean visibility.

\subsection{The PDF of Visibility}
\label{sec::V_PDF}

We now derive the PDF of visibility, when the data explore a representative ensemble of the diffractive scintillation. This result relies on both the PDF of visibility within each scintillation snapshot (\S\ref{sec::Snapshot_Approx}) and the PDF of the scintillation random variables $\{ z_{\rm g}, z_{\rm g}'\}$:
\begin{align}
P(V;N)=\int P(V;N|z_{\rm g}, z_{\rm g}') P(z_{\rm g},z_{\rm g}') d^2z_{\rm g} d^2 z_{\rm g}'.
\end{align}
Now, $z_{\rm g}$ and $z_{\rm g}'$ are drawn from a distribution of circular complex Gaussian random variables with some correlation \mbox{$\rho_{\rm g} \equiv \langle z_{\rm g} z_{\rm g}'^\ast \rangle_{\rm S}$}. Note that we use the subscripted brackets $\langle \hdots \rangle_{\rm S}$ to designate an ensemble average over the scintillation.

Although a single realization of the scintillation pattern has a complex mean visibility $z_{\rm g} z_{\rm g}'^\ast\!$, we assume that the ensemble-averaged mean visibility $ \rho_{\rm g} \equiv \left \langle z_{\rm g} z_{\rm g}'^\ast \right \rangle_{\rm S}$ is real. In practice, this assumption merely reflects an appropriate added, constant phase to the delay model.  
This mean visibility depends on the baseline, scattered image size, and observing wavelength, and can be expressed in terms of the phase structure function $D_\phi(\textbf{b})$ of the scattering medium \citep{Tatarskii_1971,Lee_Jokipii_1,Rickett90}:
\begin{align}
\rho_{\rm g} = \exp\left[ -\frac{1}{2} D_{\phi}(\textbf{b}) \right].
\end{align}

The joint PDF $P(z_{\rm g},z_{\rm g}')$, the standardized bivariate complex Gaussian distribution, follows easily from a distribution of four correlated (real) Gaussian random variables that correspond to the real and imaginary parts of $z_{\rm g}$ and $z_{\rm g}'$; see \citet{Goodman_Complex} or \citet{IVSS}, for example. 
In terms of the scintillation norms, $r \equiv |z_{\rm g}|$ and $r' \equiv |z_{\rm g}'|$, and their relative phase $\theta \equiv \mathrm{arg}\left(z_{\rm g} z_{\rm g}'^\ast  \right)$, we have
\begin{align}
\label{eq::BV_Polar}
P(r,r'\!,\theta) &= \frac{2}{\pi} \frac{r r'}{\left(1 - \rho_{\rm g}^2 \right)} \exp\left[ -\frac{\left( r^2 + r'^2 - 2\rho_{\rm g} r r' \cos \theta \right)}{\left(1 - \rho_{\rm g}^2 \right)}  \right].
\end{align}

\subsubsection{The i.i.d.\ Approximation}

\begin{figure}[t]
\includegraphics*[width=0.95\textwidth]{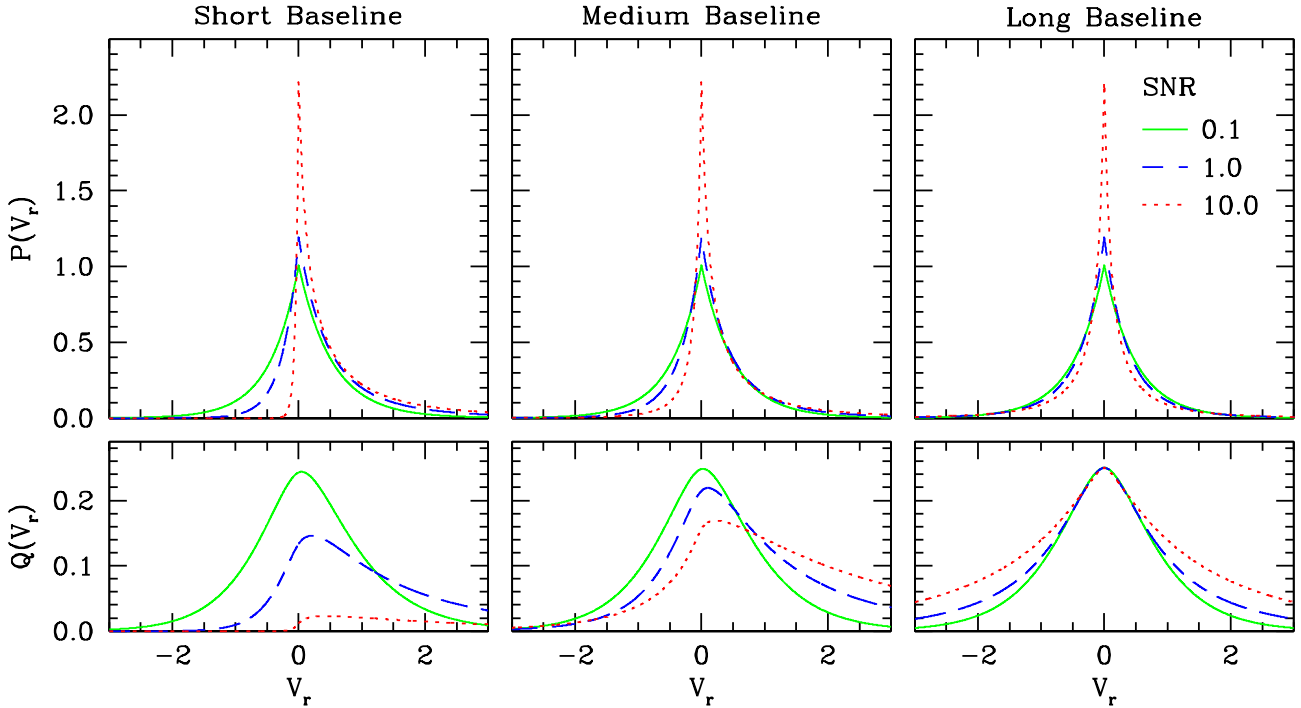}
\caption{
Visibility projections as a function of baseline and signal-to-noise (SNR) for $N=1$. We have set the pulse amplitude $A$ to be unity, the receivers to have identical signal-to-noise ratios ($I_{\rm s} = I_{\rm s}'$ and $I_{\rm n} = I_{\rm n}'$), and the single-dish intensities to be fixed: $I_{\rm s} + I_{\rm n} = 1$. The source contribution can \emph{increase} the variance about the real axis on long baselines as a result of scintillation. This combination of projections demonstrates that, even without averaging, a strong scintillating source may be readily detected regardless of baseline. The short, medium, and long baselines correspond to $\rho_{\rm g}=1.0$, $0.5$, and $0.0$, respectively.
}
\label{fig::SNR_N=1}
\end{figure}

For the i.i.d.\ approximation (\S\ref{sec:iid}), we can further reduce the expression for $P(V;N)$ by integrating $\theta$ because $\mathrm{arg}(\rho)$ is the only quantity in Eq.\ \ref{eq::V_noscint} that depends on $\theta$. We then obtain
\begin{align}
\label{V_PDF_reduced}
P(V;N) = \frac{N^{N+1}}{2^{N-2} \pi (N-1)!} \frac{|V|^{N-1}}{\left( 1 - \rho_{\rm g}^2 \right)} \int_0^\infty dr dr'\, &\frac{\left( 1 - |\rho|^2 \right)^N}{a^{N+1}} 
\text{K}_{N-1}\left( N \frac{|V|}{a} \right) r r' \exp\left(- \frac{r^2 + r'^2}{1-\rho_{\rm g}^2} \right) \\
\nonumber & \qquad {} \times I_0 \left( \sqrt{  \left(\frac{2 \rho_{\rm g} r r'}{\left( 1 - \rho_{\rm g}^2 \right)} \right)^2  + \left(\frac{N|\rho V|}{a} \right)^2 +  \frac{ 4N\rho_{\rm g}}{1-\rho_{\rm g}^2} |\rho| r r' \frac{\mathrm{Re}(V)}{a} }  \right).
\end{align}
In this expression, $I_0(x)$ is the modified Bessel function of the first kind. The visibility projections and their generalizations, such as the projected imaginary variance $\int dV_{\rm i}\, V_{\rm i}^2 P(V)$, similarly follow from the results of \S\ref{sec::Visibility_Projections}. Figure \ref{fig::SNR_N=1} illustrates the effects of baseline and signal-to-noise on the projections.

\subsubsection{The Gaussian\ Approximation}
\label{sec::Gaussian_Approx_wScint}
We now include the effects of scintillation for the Gaussian approximation (\S\ref{sec::Gaussian_Approximation_V}). The distribution of visibility, after including the scintillation ensemble but before including the effects of noise, can be written (see Eq.\ \ref{eq::V})
\begin{align}
\label{eq::V_Neqinf}
P(V;N\rightarrow \infty ) &= \frac{2}{\pi \kappa^2} \frac{1}{\left(1-\rho_{\rm g}^2\right)} K_0 \left( \frac{2}{\left(1 - \rho_{\rm g}^2 \right)} \frac{|V|}{\kappa} \right)
\exp\left( \frac{2\rho_{\rm g}}{\left(1-\rho_{\rm g}^2 \right)} \frac{\mathrm{Re}[V]}{\kappa} \right),
\end{align}
where $\kappa \equiv \langle A \rangle_N \sqrt{I_{\mathrm{s}}I_{\mathrm{s}}'}$ and $\rho_{\rm g} \equiv \left \langle z_{\rm g} z_{\rm g}'^\ast \right \rangle$. 
\citet{IVSS} described this limit and result.

As noted in \S\ref{sec::Gaussian_Approximation_V}, the noise only depends on the mean visibility when the baseline is short (relative to the diffractive scale). Hence, the Gaussian approximation, including the scintillation ensemble, becomes
\begin{align}
\label{eq::VGwithNoise}
P(V;N ) &\approx \int P(V';N\rightarrow \infty ) P_{\rm noise}(V - V',V') d^2V'.
\end{align}
In this case, $P_{\rm noise}(V, V_0)$ denotes the elliptical Gaussian distribution of noise centered on $V_0$. The orientation of the ellipse is given by the phase of $V_0$; the major and minor axes depend on both $V_0$ and the noise coefficients $\{b_0, b_1, b_2\}$, as derived in \S\ref{sec::Snapshot_Moments}.

From Eq.\ \ref{eq::VGwithNoise}, we see that the visibility projections after including the effects of noise can be applied directly to $P_{\rm noise}$. For example, the real projection is given by
\begin{align}
P(V_{\rm r};N ) &\approx \int P(V';N\rightarrow \infty ) \left\{ \frac{1}{\sqrt{2\pi}} \frac{1}{\sqrt{b_0 + b_1 |V'| + b_2|V'|^2\cos^2\phi}}\exp\left[\frac{-\left(V_{\rm r} - V'_{\rm r} \right)^2}{2\left(b_0 + b_1 |V'| + b_2|V'|^2\cos^2\phi\right)}\right] \right\} d^2V',
\end{align}
where $\phi \equiv \mathrm{arg}\ V'$. \citet{Vela_noise} and \citet{Vela_18cm} used these representations to characterize the PDF of visibility for observations of the Vela pulsar.

\begin{figure}[t]
\centering
\includegraphics*[width=0.6\textwidth]{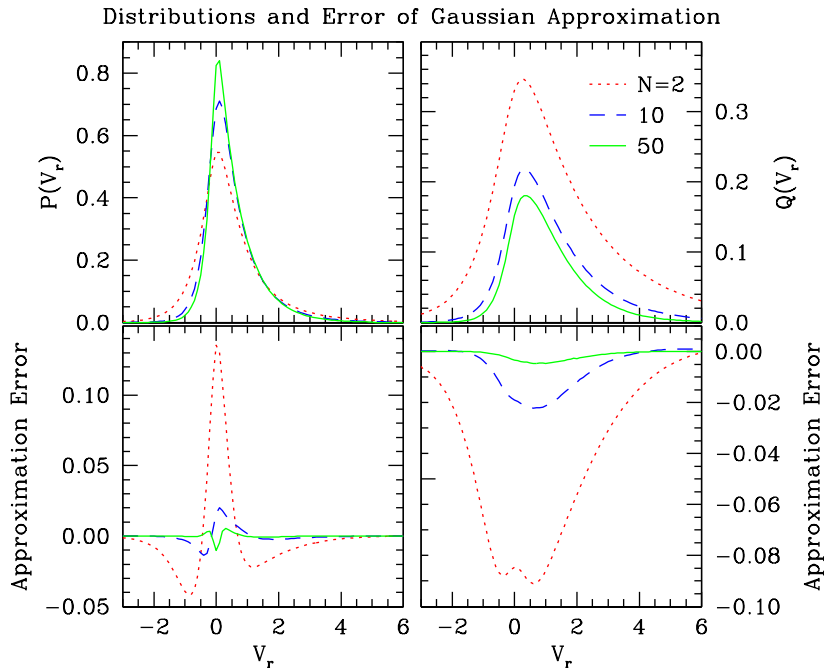}
\caption{
Visibility projections $P(V_{\rm r})$ and $Q(V_{\rm r})$ for $N=2$, 10, and 50, and the corresponding errors $\delta P$ and $\delta Q$ in the Gaussian approximation. We have set the signal-to-noise ratio to unity ($I_{\rm s} = I_{\rm s}' = I_{\rm n} = I_{\rm n}' = 1)$ and assigned a moderate baseline ($\rho_{\rm g} = 0.5$). We have set all pulse amplitudes to be equal to one, so the i.i.d.\ approximation is exact. 
}
\label{fig::Gaussian_Approx}
\end{figure}

The Gaussian approximation is also effective on longer baselines, although the noise depends on the scintillation pair $\{z_{\rm g},z_{\rm g}'\}$ rather than just the mean visibility. However, in this case, we simply replace $b_1$ by its value when $z_{\rm g} = z_{\rm g}'$; in this case, the noise only depends on the mean visibility, and Eq.\ \ref{eq::VGwithNoise} applies. Figure \ref{fig::Gaussian_Approx} shows the error in the Gaussian approximation for various degrees of averaging on a moderate baseline.

\subsection{Examples}
\label{sec::Examples}

Because the form of the visibility PDF is rather opaque, we now refine it for several cases of interest. The zero-baseline interferometer (\S\ref{sec:Ex_Zero_Baseline}) offers the most substantial reduction of complexity and relates neatly to the analogous results for intensity. At the opposite extreme, an infinite baseline (\S\ref{sec::Ex_Long_Baseline}) also leads to a greatly simplified form, albeit with no fundamental decrease in numerical difficulty. Finally, the regime of high signal-to-noise (\S\ref{sec::Ex_High_SNR}) highlights the influence of self-noise on the PDF and lays the foundation for calculating the asymptotic ($|V|\rightarrow \infty$) dependence (see \S\ref{sec::Asymptotic}). For simplicity, we derive all results of this section using the i.i.d.\ approximation.

\subsubsection{Zero-Baseline}
\label{sec:Ex_Zero_Baseline}
The zero-baseline limit for visibility is particularly simple: $\rho_{\rm g} \rightarrow 1$. In this case, $z_{\rm g} = z_{\rm g}'$, and the effects of scattering depend only on the scintillation ``gain'' $G \equiv |z_{\rm g}|^2$:
\begin{align}
\label{eq::zero_baseline}
P(V;N) = \frac{N^{N+1}}{2^N \pi (N-1)!} |V|^{N-1} \int_0^\infty dG\, &\frac{\left( 1 - |\rho|^2 \right)^N }{a^{N+1}} \text{K}_{N-1}\left( N \frac{|V|}{a} \right) \exp \left( N \frac{\mathrm{Re}\left[V \rho^\ast \right]}{a} \right) P(G),
\end{align}
where $P(G) = e^{-G}$. This representation has two distinct advantages: it requires a single integral, and it can be easily modified to any alternate distribution of scintillation gain, as might occur from an extended emission region or weak scattering, for example.

We can also use the results of \S\ref{sec::Visibility_Projections} to evaluate projections of the visibility distribution without requiring an additional numerical integral:
\begin{align}
P(V_{\rm r};N) &= \frac{N}{(N-1)!} \int_0^\infty dG\, \frac{1}{a}   \left[ \frac{N}{2}\left(1-\rho^2 \right) \frac{|V_{\rm r}|}{a} \right]^{N} k_{N-1}\left( N  \frac{|V_{\rm r}|}{a} \right) e^{N \rho \frac{V_{\rm r}}{a}} P(G),\\
\nonumber Q(V_{\rm r};N) &= \frac{2}{N!} \int_0^\infty dG\, \frac{a}{\left( 1 - \rho^2 \right)} \left[ \frac{N}{2} \left( 1 - \rho^2 \right) \frac{|V_{\rm r}|}{a} \right]^{N+1}\! k_{N}\left( N \frac{|V_{\rm r}|}{a} \right) e^{N \rho \frac{V_{\rm r}}{a}} P(G).
\end{align}
In these expressions, $k_N(x)$ is the modified spherical Bessel function of the second kind \citep{Arfken}. 

As the averaging increases, the distribution of visibility approaches $P(G)$:
\begin{align}
P(V;N\rightarrow \infty) = \frac{1}{\langle V \rangle} e^{-\mathrm{Re}[V]/\langle V \rangle} \theta( \mathrm{Re}[V]) \delta( \mathrm{Im}[V] ),
\end{align}
where $\theta(x)$ is the Heaviside function, and $\delta(x)$ is the Dirac delta function.

\subsubsection{Long-Baseline}
\label{sec::Ex_Long_Baseline}
On a sufficiently long baseline, the interferometer completely resolves the scattering disk, and the respective propagation kernels at the two stations become completely independent: $\rho_{\rm g} \rightarrow 0$. 
Although ground-based VLBI can only marginally achieve this regime for a few of the most heavily scattered pulsars at meter and decimeter wavelengths \citep[e.g.][]{Gwinn_93}, space VLBI with RadioAstron can easily resolve the scattering disks of many pulsars \citep{Kardashev_2009}. A precise understanding of the visibility statistics for ultra-long baselines will help to maximize the information that can be gleaned from space VLBI. 

In this limit, Eq.\ \ref{V_PDF_reduced} becomes
\begin{align}
\label{V_PDF_reduced_longbaseline}
&P(V;N) = \frac{N^{N+1}}{2^{N-2} \pi (N-1)!} |V|^{N-1} \int_0^\infty dr dr'\, \frac{\left( 1 - |\rho|^2 \right)^N}{a^{N+1}} \text{K}_{N-1}\left( N \frac{|V|}{a} \right) I_0 \left( N |\rho| \frac{|V|}{a}  \right) r r' e^{- \left(r^2 + r'^2\right)}.
\end{align}
Observe that this distribution depends only on $|V|$, as necessitated by the phase invariance of this limit. 
Although the mean visibility is zero, the effects of scintillation can still be substantial, as Figure \ref{fig::SNR_N=1} illustrates.

\subsubsection{High Signal-to-Noise}
\label{sec::Ex_High_SNR}

In the limit of infinite signal-to-noise, the fields at two stations will exhibit identical noise, multiplied by their respective complex scintillation gains. 
In terms of the field statistics (\S\ref{sec::Field_Statistics}), each visibility is the average of $N$ exponential random variables with means $A_j \sqrt{I_{\rm s} I_{\rm s}'} z_{\rm g} z_{\rm g}'^\ast$. 

For simplicity, we apply the i.i.d.\ approximation. In this case, each averaged visibility is drawn from an Erlang distribution, multiplied by the scintillation gain, which rotates the phase to $\mathrm{arg}\left( z_{\rm g} z_{\rm g}'^\ast \right)$. Expressed in terms of the magnitude and phase of the visibility, $V=|V| e^{i\phi}$, this distribution is
\begin{align}
P\left(|V|,\phi;N\Big|z_{\rm g},z_{\rm g}'\right) = \frac{N^N}{(N-1)!} \frac{|V|^{N-1}}{\left( A \sqrt{I_{\rm s} I_{\rm s}'} r r' \right)^{N}} \exp\left[ -N \frac{|V|}{A \sqrt{I_{\rm s} I_{\rm s}'} r r' } \right] \delta\left( \phi - \theta \right).
\end{align}
The definitions of $r$, $r'$, and $\theta$ are equivalent to those in \S\ref{sec:iid}. When combined with the scintillation ensemble, $\theta$ can be trivially integrated against its delta function to give
\begin{align}
\label{eq::SND}
\nonumber P\left(|V|,\phi;N\right) &= 
\frac{2}{\pi} \frac{N^N}{(N-1)!}  \frac{1}{\left(1 - \rho_{\rm g}^2 \right)} \frac{|V|^{N-1}}{\left( A \sqrt{I_{\rm s} I_{\rm s}'} \right)^{N}} 
\int_0^\infty dr dr'\,
\frac{1}{\left(r r' \right)^{N-1}} \exp\left[ -\frac{N |V|}{A \sqrt{I_{\rm s} I_{\rm s}'} r r' }  -\frac{\left( r^2 + r'^2 - 2\rho_{\rm g} r r' \cos \phi \right)}{\left(1 - \rho_{\rm g}^2 \right)}  \right]\\
\nonumber &=
\frac{2N}{\pi(N-1)!}  \frac{1}{\left(1 - \rho_{\rm g}^2 \right)} \frac{1}{ A \sqrt{I_{\rm s} I_{\rm s}'}}
\left( \frac{2N |V|}{A \sqrt{I_{\rm s} I_{\rm s}'}} \right)^{\!\frac{N}{2}}\!\!
\int_0^{\pi/2} d\psi\,
\frac{1}{\beta(\psi)} \left[ \frac{\beta(\psi)}{\sin(2\psi)} \right]^{\frac{N}{2}}\!
K_{N-2} \left(2\sqrt{2 N \frac{\beta(\psi)}{\sin(2\psi)}  \frac{|V|}{ A\sqrt{I_{\rm s} I_{\rm s}'}}   } \right),\\
\beta(\psi) &\equiv \frac{1 - \rho_{\rm g} \cos \phi \sin(2\psi)}{1 - \rho_{\rm g}^2}.
\end{align}
To obtain the second line, we transformed to polar coordinates $\{r = \ell \cos \psi,\, r' = \ell \sin \psi\}$ and integrated over $\ell$. Thus, as for the zero-baseline case, we have reduced the integration to a single dimension. We next use this form to determine the asymptotic behavior of $P(V)$.

\subsection{Asymptotic Behavior}
\label{sec::Asymptotic}

We now derive the behavior of $P(V;N)$ as $|V| \rightarrow \infty$. To proceed, we first consider the high signal-to-noise results. At large $|V|$, the Bessel function of Eq.\ \ref{eq::SND} approaches an exponential, so we can apply the method of steepest descent to approximate the integral over $\psi$ \citep{Arfken}: 
\begin{align}
\label{eq::asymptotic}
P\left(|V|\rightarrow \infty,\phi;N\right) \propto |V|^{\frac{N-1}{2}} 
\exp\left[ -2\sqrt{ 2N \left( \frac{1 - \rho_{\rm g}\cos\phi}{1-\rho_{\rm g}^2} \right) \frac{|V|}{A \sqrt{I_{\rm s} I_{\rm s}'}} }\ \right].
\end{align}
Here, the constant of proportionality is also a function of $\phi$.

On the other hand, the distribution of visibility for purely background noise has the asymptotic form (see Eq.\ \ref{eq::V_Niid})
\begin{align}
P\left(|V|\rightarrow \infty,\phi;N,I_{\rm s} = I_{\rm s}' = 0\right) \propto |V|^N K_{N-1} \left( \frac{2 N |V|}{\sqrt{I_{\rm n} I_{\rm n}'} } \right)
\sim |V|^{N-1/2} \exp \left( -\frac{2 N |V|}{\sqrt{I_{\rm n} I_{\rm n}'}} \right).
\end{align}
Thus, the stronger, scintillation-induced ``skirt'' of source power will dominate the PDF of visibility in the asymptotic regime defined by 
\begin{align}
|V| \gg \frac{2}{N}\left( \frac{1-\rho_{\rm g} \cos \phi}{1-\rho_{\rm g}^2} \right) \frac{I_{\rm n} I_{\rm n}'}{A\sqrt{I_{\rm s} I_{\rm s}'}}.
\end{align}

\begin{figure}[t]
\centering
\includegraphics*[width=0.9\textwidth]{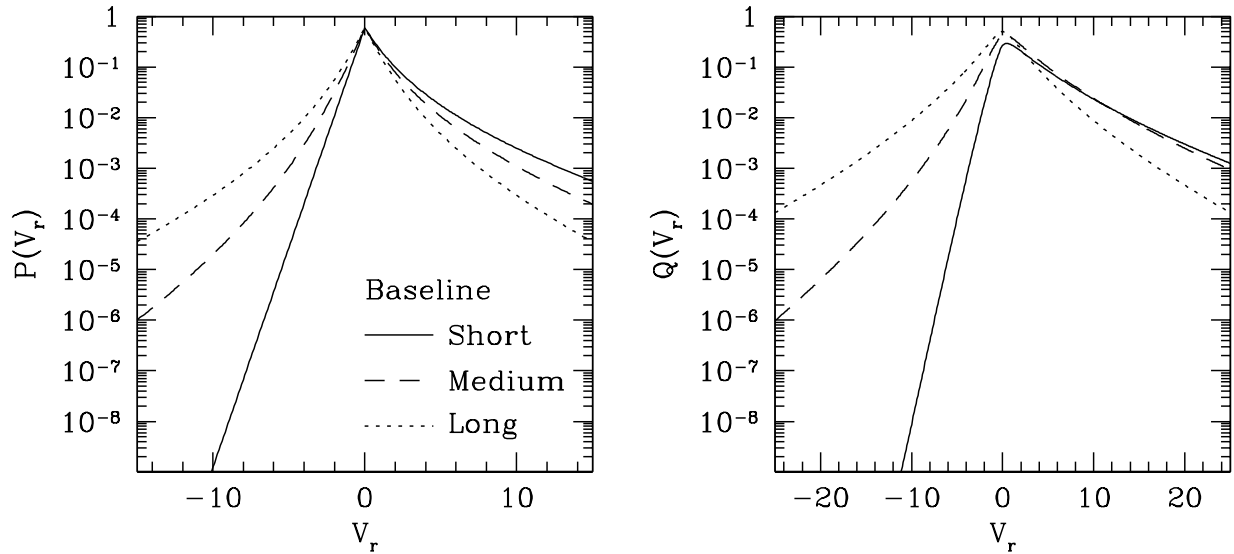}
\caption{
Exact projections $P(V_{\rm r})$ and $Q(V_{\rm r})$ for $N=1$; the source and background amplitudes are unity. The broad scintillation-induced wings are distinctive at large $|V_{\rm r}|$. However, for zero-baseline and negative $V_{\rm r}$, there is only the exponentially-falling noise contribution, reflecting the absence of source power on that half-line and, mathematically, the singularities in Eqs.\ \ref{eq::asymptotic} \& \ref{eq::Asymptotic_Projections}.
}
\label{fig::Asymptotic}
\end{figure}

Of course, the projections also reflect this remarkable behavior. For example, the projection onto $V_{\rm r} \equiv \mathrm{Re}(V)$ follows by applying the method of steepest descent to these asymptotic forms. We thereby obtain
\begin{align}
\label{eq::Asymptotic_Projections}
P\left(V_{\rm r}\rightarrow \pm\infty;N,I_{\rm s}, I_{\rm s}' > 0\right) &\propto |V_{\rm r}|^{\frac{N}{2}-\frac{3}{4}}\exp\left[ -2\sqrt{ 2N \left( \frac{1}{1+\mathrm{sign}(V_{\rm r})\rho_{\rm g}} \right) \frac{|V_{\rm r}|}{A \sqrt{I_{\rm s} I_{\rm s}'}} }\ \right],\\
\nonumber P\left(V_{\rm r}\rightarrow \pm\infty;N,I_{\rm s} = I_{\rm s}' = 0\right) &\propto |V_{\rm r}|^{N-1} \exp\left( -\frac{2N |V_{\rm r}|}{\sqrt{I_{\rm n} I_{\rm n}'}} \right).
\end{align}
Once again, a scintillating source extends the wings and introduces a baseline-dependent asymmetry. In this case, the constant of proportionality is a function of $\mathrm{sign}(V_{\rm r})$, and the asymptotic regime is determined by
\begin{align}
|V_{\rm r}| \gg \frac{2}{N} \frac{1}{\left( 1+\mathrm{sign}(V_{\rm r})\rho_{\rm g} \right)} \frac{I_{\rm n} I_{\rm n}'}{A\sqrt{I_{\rm s} I_{\rm s}'}}.
\end{align}
Figure \ref{fig::Asymptotic} demonstrates this asymptotic behavior of the projections on various baselines.

Observe that both these asymptotic forms have poles for $\rho_{\rm g} = \pm 1$. This behavior reflects the fact that, in these cases, the signal power is restricted to the real half-line that matches the sign of the mean visibility, so the asymptotic behavior elsewhere corresponds to that of pure background noise.

Perhaps most surprisingly, in the limit $N\rightarrow \infty$ (given by Eq.\ \ref{eq::V_Neqinf}), the asymptotic behavior is exponential, regardless of the signal-to-noise. Hence, the broad $\exp(-\sqrt{|V|})$ skirt arises from the delicate interplay of the scintillation and the self-noise.

\subsection{Estimation of the Self-Noise}
\label{sec::VSelfNoise}

As a last application of these visibility statistics, we now present a prescription that can be applied to quantify self-noise. 
As we have discussed, self-noise provides a powerful diagnostic of a signal, particularly when characterizing intrinsic variability. For example, source variability on timescales shorter than $t_{\rm acc}$ induces correlations in the spectral noise, without modifying the mean spectrum \citep{Intermittent_Noiselike_Emission}. We now outline one technique that estimates self-noise and thereby facilitates a detection of such correlations.

Our method is similar to its analog for intensity \citepalias[\S5.2]{IPDF}. Namely, we use pairs of nearby samples to calculate finite estimates of the signal and noise. Each pair of samples is assumed to be within a single scintillation element. We will assume that the pair consists of samples with uncorrelated self-noise (e.g.\ pairs from different pulses, or from the same pulse with negligible intrinsic modulation). We calculate the noise in pairs of samples as a function of their mean. However, because the visibility is complex, the noise must be treated as a vector quantity as it has different behavior in phase and at quadrature with the signal (see \S\ref{sec::Snapshot_Moments} and Figure \ref{fig::V_Noise}). We find it advantageous to work with analytical expressions of the involved complex quantities and, thus, define
\begin{align}
\delta V^2\!\left(V \right) \equiv 2 \left \langle \left(V_1 - \frac{V_1 + V_2}{2} \right)^2 \right \rangle,\quad V \equiv \frac{V_1 + V_2}{2}.
\end{align}
We again work within the i.i.d.\ approximation. The distribution of visibility is then given by Eq.\ \ref{eq::V_Niid}, and we obtain,
\begin{align}
\label{eq::V_Self_Noise}
\delta V^2\!\left(V\right) = \frac{V^2}{N + \frac{1}{2}}.
\end{align}
As for intensity, this method for estimating the self-noise agrees with the exact expression with $N \rightarrow N + 1/2$.

If the averaged visibilities are from different pulses, then pulse-to-pulse variations can contribute additional noise. If the averaged visibilities are from the same pulses, then intrinsic variations on timescales shorter than $t_{\mathrm{acc}}$ induce correlations in self-noise, and thereby decrease the measured noise; if $N>1$ then pulse-to-pulse variations within the averaging will increase the noise. 
\citet{Noise_Inventory} applied similar tests to infer short-timescale variability ($<300 \mu \mathrm{s}$) of PSR B0834+06.

\section{Effects of an Extended Emission Region on the Visibility PDF}
\label{sec::PDF_Emission_Size}

We now derive the modifications to the PDF of visibility from a spatially-extended emission region. Such emission superimposes many slightly-offset copies of the diffraction pattern at the observer, suppressing the observed scintillation. Optical scintillation provides a familiar demonstration: ``Stars twinkle, but planets do not.'' Emission extending over a region much larger than the diffractive scale $r_{\rm d}$ (see \S\ref{sec::obs_strategy}) quenches the scintillation, and so diffractive scintillation studies can effectively probe emission scales $\lsim r_{\rm d}$.

\subsection{The Effects of an Extended Emission Region on the Field Statistics}

If the emission spans a transverse size $\ll \! r_{\rm d}$, then the modification to the field statistics can be derived quite generally. Explicitly, in terms of transverse source coordinates $\textbf{s}$ and the notation of \S\ref{sec::Field_Statistics}, the observed electric field takes the form
\begin{align}
\label{eq::size_V_1}
\tilde{x}_i = \left\{ \int d^2 \textbf{s} \ \sqrt{A(\textbf{s}) I_{\mathrm{s}}} z_{\mathrm{f}}(\textbf{s}) z_{\mathrm{g}}(\textbf{s}) \right\} + \sqrt{I_{\mathrm{n}}} z_{\mathrm{b}}.
\end{align}
We set the origin of the coordinates $\textbf{s}$ so that $\int d^2\textbf{s}\ \textbf{s} A(\mathbf{s}) = \mathbf{0}$. Because the source intensity $I(\textbf{s})$ is assumed to be confined within a region $\ll\! r_{\rm d}$, $z_{\mathrm{g}}(\textbf{s})$ will only vary slightly and we may expand to linear order: $z_{\mathrm{g}}(\textbf{s}) \approx z_{\mathrm{g}}(\textbf{0}) + \left. (\textbf{s} \cdot \nabla)  z_{\mathrm{g}} \right \rfloor_{s=0}$.

The source term in Eq.\ \ref{eq::size_V_1} is then a convolution of three complex Gaussian random variables. Because of our choice of origin for $\textbf{s}$, at linear order these three random variables are mutually uncorrelated during a fixed scintillation pattern. In addition, the scintillation random variable, $z_{\rm g}(\textbf{s})$, is uncorrelated with its spatial derivatives, so the scales of the three respective variances are also mutually independent at linear order. 
Combining these characteristics for the pair of receivers then gives the form of the visibility field statistics to leading order:
\begin{align}
\label{eq::V_withsize}
V \approx \frac{1}{N} \frac{1}{1 + \gamma_{\mathrm{s},1} + \gamma_{\mathrm{s},2}} \sum_{j=1}^N & \left[ \sqrt{A_j I_{\mathrm{s}}} \left( z_{\mathrm{f},j} z_{\rm g} + \sqrt{\gamma_{\mathrm{s},1}} z_{\mathrm{f},1,j} z_{\mathrm{g},1} + \sqrt{\gamma_{\mathrm{s},2}} z_{\mathrm{f},2,j} z_{\mathrm{g},2} \right) + 
\sqrt{I_{\mathrm{n}}} z_{\mathrm{b},j} \right]\\
&\nonumber \qquad {} \times \left[ \sqrt{A_j I_{\mathrm{s}}'} \left( z_{\mathrm{f},j} z_{\rm g}' + \sqrt{\gamma_{\mathrm{s},1}} z_{\mathrm{f},1,j} z_{\mathrm{g},1}' + \sqrt{\gamma_{\mathrm{s},2}} z_{\mathrm{f},2,j} z_{\mathrm{g},2}' \right) + \sqrt{I_{\mathrm{n}}'} z_{\mathrm{b},j}' \right]^\ast.
\end{align}
Here, the paired scintillation random variables are correlated, $\rho_{\rm g} \equiv \langle z_{\rm g} z_{\rm g}'^\ast \rangle$ and  $\rho_{{\rm g},i} \equiv \langle z_{{\rm g},i} z_{{\rm g},i}'^\ast \rangle$, but all other pairs of random variables $\{ z_x, z_y \}$ are uncorrelated. We have chosen the scaling prefactor so that the intensity measured by either receiver is unaffected by the extent of the emission. Our expansion parameters, the dimensionless subsidiary scales $\gamma_{\mathrm{s},i} \ll 1$, contain information about the transverse extent of the source emission. More specifically, these scales are proportional to the spatial standard deviation of \emph{integrated} flux density. For example, spatially-offset, pointlike emission sites need only to emit within the same accumulation time, but not necessarily at the same retarded time, to affect the scintillation statistics. Hence, even for emission that is highly beamed and, thus, instantaneously pointlike, this method can identify a transverse size that relates to the emission altitude; see \citet{GUPPI_Size_2}.

Within a fixed scintillation pattern, Eq.\ \ref{eq::V_withsize} is the $N$-fold convolution of products of complex circular Gaussian random variables, as for a point source. Thus, applying the i.i.d.\ approximation, we see that the PDF of visibility within each scintillation snapshot takes the same form as Eq.\ \ref{eq::V_noscint}. However, the correlation of each multiplied pair depends on the extended emission region; the subsequent inclusion of the scintillation ensemble is complicated by the different correlations for each scintillation factor: $\rho_{\rm g} \neq \rho_{\mathrm{g},1} \neq \rho_{\mathrm{g},2}$. The appropriate correlation and respective intensities of the multiplied terms are
\begin{align}
\label{size_correlation}
\rho &= \left( \frac{ z_{\mathrm{g}} z_{\mathrm{g}}'^\ast + \gamma_{\mathrm{s},1} z_{\mathrm{g},1} z_{\mathrm{g},1}'^\ast + \gamma_{\mathrm{s},2} z_{\mathrm{g},2} z_{\mathrm{g},2}'^\ast }{ 1 + \gamma_{\mathrm{s},1} + \gamma_{\mathrm{s},2} } \right) A \sqrt{\frac{I_{\rm s} I_{\rm s}'}{ \bar{I} \bar{I}'}},\\
\nonumber \bar{I} &\equiv A I_{\rm s}\left( \frac{ |z_{\rm g}|^2 + \gamma_{\mathrm{s},1} |z_{\mathrm{g},1}|^2+ \gamma_{\mathrm{s},2} |z_{\mathrm{g},2}|^2 }{ 1 + \gamma_{\mathrm{s},1} + \gamma_{\mathrm{s},2} }  \right) + I_{\rm n},\\
\nonumber \bar{I}' &\equiv A I_{\rm s}'\left( \frac{ |z_{\rm g}'|^2 + \gamma_{\mathrm{s},1} |z_{\mathrm{g},1}'|^2+ \gamma_{\mathrm{s},2} |z_{\mathrm{g},2}'|^2 }{ 1 + \gamma_{\mathrm{s},1} + \gamma_{\mathrm{s},2} }  \right) + I_{\rm n}'.
\end{align}

Also, observe that the mean visibility is weakly diminished by a finite source emission size:
\begin{align}
\left \langle V \right \rangle \equiv \frac{\rho_{\rm g} + \gamma_{\mathrm{s},1} \rho_{\mathrm{g},1} + \gamma_{\mathrm{s},2} \rho_{\mathrm{g},2} }{1+\gamma_{\mathrm{s},1} + \gamma_{\mathrm{s},2}} \leq \left \langle V \right \rangle_{\gamma_{\mathrm{s},i}=0}.
\end{align}

\subsection{ The Relation Between the Emission Region and the Dimensionless Size Parameters $\gamma_{{\rm s},i}$ }

The precise correspondence between the source dimensions and $\gamma_{{\rm s},i}$ requires knowledge of the distribution of scattering material. The relationship between the correlations $\{\rho_{\rm g},\rho_{{\rm g},i}\}$ and the observing baseline likewise depends on the scattering assumptions. For example, by assuming a square-law phase structure function, \citet{IVSS} derived the relations
\begin{align}
\gamma_{\mathrm{s},i} = \left( \frac{D}{R} k \theta_i \sigma_i \right)^2\!,\qquad
\rho_{\mathrm{g},i} \approx \left[ 1 - \left(b_i k \theta\right)^2 \right] \exp\left[ -\frac{1}{2} \left( |\textbf{b}| k \theta \right)^2 \right] =  \left[ 1 - \left(b_i k \theta\right)^2 \right]  \rho_{\rm g}.
\end{align}
Here, $D$ is the characteristic observer-scatterer distance, $R$ is the characteristic source-scatterer distance, $k$ is the observing wavenumber, $\theta_i$ is the angular size of the scattering disk along $\hat{s}_i$, and $\sigma_i$ is the standard deviation of the (integrated) distribution of source intensity along $\hat{s}_i$. The scintillation correlations also depend on the baseline length $|\textbf{b}|$ and its projections $b_i$ along $\hat{s}_i$. Consequently, for baselines much shorter than the diffractive scale, the three correlations are nearly equal. 
Thus, the dimensionless size parameters $\gamma_{\mathrm{s},i}$ give the squared size of the source in orthogonal directions $\hat{s}_i$, in units of the magnified diffractive scale.

\subsection{Approximate Evaluation of the PDF of Visibility}

We now apply the i.i.d.\ approximation to estimate the PDF of visibility, including the effects of an extended emission region. Namely, we combine the form of the PDF of visibility within each scintillation snapshot, given by Eq.\ \ref{eq::V_noscint}, with the distribution of each scintillation random variable:
\begin{align}
\label{eq::V_size}
P(V;N,\gamma_{\mathrm{s},1},\gamma_{\mathrm{s},2}) = \frac{N^{N+1}}{2^N \pi (N-1)!}  |V|^{N-1}
\int d^3\textbf{r} d^3\textbf{r}_1 d^3\textbf{r}_2 
& \frac{\left( 1 - |\rho|^2 \right)^N }{a^{N+1}} 
\text{K}_{N-1}\left( N \frac{|V|}{a} \right) \exp \left( N \frac{\mathrm{Re}\left[V \rho^\ast \right]}{a} \right)\\
\nonumber & \qquad {} \times P(r,r',\theta) P(r_1, r_1', \theta_1) P(r_2, r_2', \theta_2).
\end{align}
Here, we have shifted to polar coordinates for each pair of scintillation gains; e.g. $\textbf{r} \equiv \{ r \equiv |z_{\rm g}|, r' \equiv |z_{\rm g}'|, \theta \equiv \arg( z_{\rm g} z_{\rm g}'^\ast) \}$. Eq.\ \ref{eq::BV_Polar} then gives the distribution $P(r,r',\theta)$ of each triplet. Also, $\rho$, $\bar{I}$, and $\bar{I}'$ are defined by Eq.\ \ref{size_correlation}, and $a$ is then as in Eq.\ \ref{eq::V_noscint}; these variables depend on the integration variables, which account for the scintillation.

While typical numerical techniques for high-dimensional integrals can evaluate Eq.\ \ref{eq::V_size}, they are computationally expensive and unenlightening. \citet{Vela_18cm} derived an efficient technique, suitable for $N\gsim 20$, using the Gaussian approximation (\S\ref{sec::Gaussian_Approximation_V}). Their method requires only a one-dimensional integral and a two-dimensional grid convolution, which accounts for the signal-dependent noise (see \S\ref{sec::Gaussian_Approx_wScint}).

However, because we are interested in small values of averaging, especially $N=1$, we present an alternative. Namely, for a small source, anisotropic emission size effects ($\gamma_{\mathrm{s},1} \neq \gamma_{\mathrm{s},2}$) only weakly modify those of an equivalent isotropic region. For the analogous effects on intensity, for example, the effects of anisotropy are quadratic in $\gamma_{\mathrm{s},i}$, and so \citetalias{IPDF} worked in terms of an equivalent isotropic size: $\gamma_{\rm s} \equiv \gamma_{\mathrm{s},1} = \gamma_{\mathrm{s},2}$. For our purposes, a one-dimensional emission region is advantageous to characterize the dominant effects of size because it obviates the integration over $P(z_{\mathrm{g},2},z_{\mathrm{g},2}')$, thereby reducing the integral of Eq.\ \ref{eq::V_size} to six dimensions.

In addition, by using the results of \S\ref{sec::Visibility_Projections}, we can evaluate projections of the visibility distribution without increasing the number of required integrations. For example, the distribution of $V_{\rm r} \equiv \mathrm{Re}(V)$ is
\begin{align}
\label{eq::V_size_projection}
P(V_{\rm r};N,\gamma_{\mathrm{s},1},\gamma_{\mathrm{s},2}) = 
\frac{N^{N+1}}{2^N(N-1)!} |V_{\rm r}|^{N} 
\int d^3\textbf{r} d^3\textbf{r}_1 d^3\textbf{r}_2 
& \frac{\sqrt{1-\rho_{\rm i}^2}}{a} \left(\frac{1-|\rho|^2 }{a \sqrt{1 - \rho_{\rm i}^2}}\right)^{N}\! k_{N-1}\left( N \sqrt{1-\rho_{\rm i}^2} \frac{|V_{\rm r}|}{a} \right) e^{N \rho_{\rm r} \frac{V_{\rm r}}{a}} \\
\nonumber & \qquad {} \times P(r,r',\theta) P(r_1, r_1', \theta_1) P(r_2, r_2', \theta_2).
\end{align}
Here, and elsewhere, the subscripts $\mathrm{r}$ and $\mathrm{i}$ denote the real and imaginary part, respectively.

\begin{figure}[t]
\centering
\includegraphics*[width=1.0\textwidth]{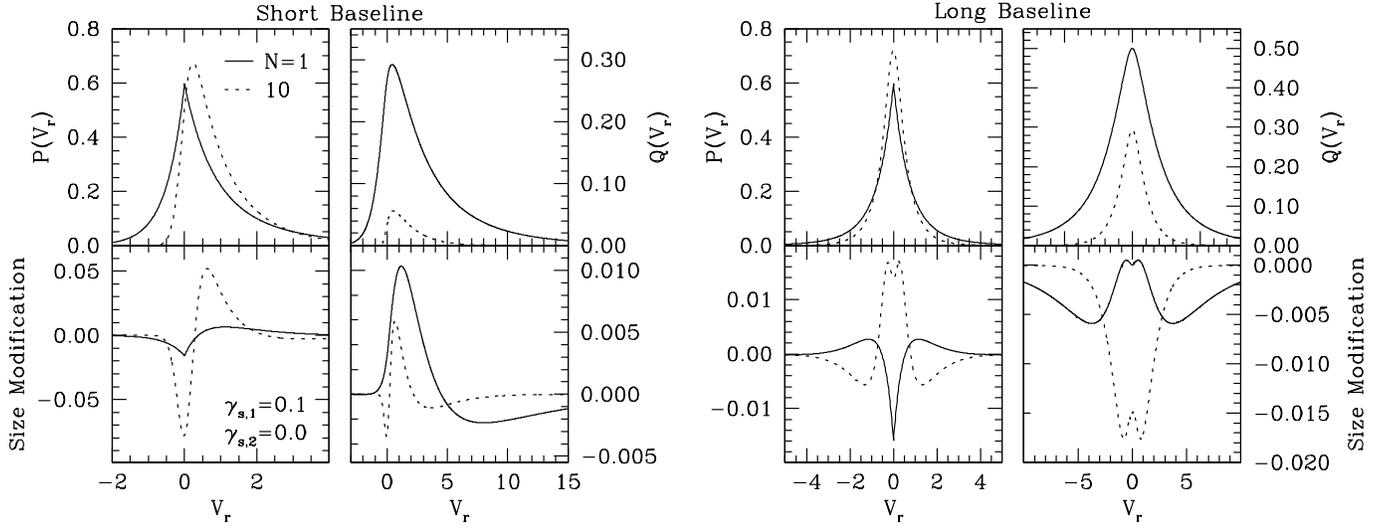}
\caption{
Visibility projections for a point source and their modifications (e.g.\ $P(V_{\rm r};\gamma_{{\rm s},i}) - P(V_{\rm r};\gamma_{{\rm s},i}=0)$) corresponding to an elongated emission region with $\gamma_{{\rm s},1}=0.1$ and $\gamma_{{\rm s},2}=0.0$. The source and background intensities are unity, as are all the pulse amplitude factors $A_j$. The modest averaging, $N=10$, reveals the W-shaped signature of emission size observed by \citet{Vela_18cm}.
}
\label{fig::Size_Effects}
\end{figure}

Figure \ref{fig::Size_Effects} shows the effects of an extended emission region on the visibility projections $P(V_{\rm r})$ and $Q(V_{\rm r})$, for both short and long baselines.

\subsection{The Short Baseline Limit}
The special case of a short baseline vastly simplifies the effects of an extended emission. In this case, the quantities in Eq.\ \ref{size_correlation} only depend on the single scintillation ``gain'' $\mathcal{G}$:
\begin{align}
\mathcal{G} \equiv \frac{ |z_{\rm g}|^2 + \gamma_{\mathrm{s},1} |z_{\mathrm{g},1}|^2+ \gamma_{\mathrm{s},2} |z_{\mathrm{g},2}|^2 }{ 1 + \gamma_{\mathrm{s},1} + \gamma_{\mathrm{s},2} }.
\end{align}
This quantity is the convolution of three independent exponential random variables. The PDF of $\mathcal{G}$ is then \citepalias[see Eq.\ A1]{IPDF}
\begin{align}
\label{eq::mult_I}
P(\mathcal{G}) = \sum_{j=1}^3 \left( \frac{\lambda_j }{\prod\limits_{\substack{\ell = 1\\ \ell\neq j}}^3 \left(\lambda_j - \lambda_\ell \right)}  \right) e^{-\mathcal{G}/\lambda_j},\quad \left\{ \lambda_1, \lambda_2, \lambda_3 \right\} \equiv \frac{1}{ 1 + \gamma_{\mathrm{s},1} + \gamma_{\mathrm{s},2} } \left\{ 1, \gamma_{\mathrm{s},1}, \gamma_{\mathrm{s},2} \right\}.
\end{align}
The PDF of visibility is then given by Eq.\ \ref{eq::zero_baseline}, with the substitution $G \rightarrow \mathcal{G}$, and the projections follow likewise. In fact, because $\rho \in \mathbb{R}$, the projections become especially tractable:
\begin{align}
\label{eq::PDF_Size_Projections}
P(V_{\rm r}) &= \frac{N}{(N-1)!} \int_0^\infty d\mathcal{G}\, \frac{1}{a}   \left[ \frac{N}{2}\left(1-\rho^2 \right) \frac{|V_{\rm r}|}{a} \right]^{N} k_{N-1}\left( N  \frac{|V_{\rm r}|}{a} \right) e^{N \rho \frac{V_{\rm r}}{a}} P(\mathcal{G}),\\
\nonumber Q(V_{\rm r}) &= \frac{2}{N!} \int_0^\infty d\mathcal{G}\, \frac{a}{\left( 1 - \rho^2 \right)} \left[ \frac{N}{2} \left( 1 - \rho^2 \right) \frac{|V_{\rm r}|}{a} \right]^{N+1}\! k_{N}\left( N \frac{|V_{\rm r}|}{a} \right) e^{N \rho \frac{V_{\rm r}}{a}} P(\mathcal{G}).
\end{align}
Figure \ref{fig::Size_Effects} illustrates how an extended emission region modifies these projections. Increased averaging tends to pronounce the effects on the real projection but not the projected imaginary variance because of the decreasing variance with averaging. Thus, $P(V_{\rm r})$ tends to be a more sensitive indicator of size than $Q(V_{\rm r})$, as noted by \citet{Vela_18cm}.

\section{Summary}
\label{sec::Summary_V}

We have derived the PDF of visibility for a strongly-scintillating source, with particular attention to spectral resolution at or near the Nyquist limit. We have incorporated background- and self-noise, source variability, the possibility of spatially-extended source emission, and arbitrary temporal averaging. We have also demonstrated that the visibility statistics exhibit several remarkable characteristics. For example, the combination of scintillation and self-noise introduces a broad ``skirt'' in the distribution of visibility that dominates asymptotic statistics, regardless of the baseline or the signal-to-noise. Finally, we have given simplifications of this PDF in various regimes, such as the zero-baseline interferometer, as well as results for various projections of the PDF.

Our results facilitate scintillation studies of pulsars in statistically-delicate regimes and studies of pulsar emission regions using interferometry. In particular, our description of Nyquist-limited statistics can provide a sensitive and robust detection of an extended emission region, which does not require any assumptions about the nature or distribution of the scattering material and can be applied to estimate the emission sizes of individual pulses.

\acknowledgments
We thank the referee for a careful reading and for several comments that improved the clarity of the text. 
We thank the U.S. National Science Foundation for financial support for this work (AST-1008865).

\appendix
\section{Mathematical Results}

\subsection{Product of Correlated Complex Gaussian Random Variables}
\label{sec::product_dist}

\label{sec::Visibility_Average}
Let $w_1$ and $w_2$ be a pair of correlated circular complex Gaussian random variables with standard deviations $\sigma_i$ and correlation $\rho \equiv \langle w_1 w_2^\ast \rangle/(\sigma_1 \sigma_2)$. \citet{IVSS} derived the PDF for the product (i.e.\ the ``visibility'') $V \equiv w_1 w_2^\ast$ for $\rho \in \mathbb{R}$. We generalize his result in two directions: by allowing $\rho \in \mathbb{C}$ and by accounting for the averaging of $N$ i.i.d.\ visibilities. The first extension is simply a complex rotation of the PDF for real $\rho$:
\begin{align}
\label{eq::V}
P(V;\sigma_1,\sigma_2,\rho) &= \frac{2}{\pi} \frac{1}{\left(1-|\rho|^2\right) \sigma_1^2 \sigma_2^2} K_0\left(\frac{2}{\left(1-|\rho|^2\right)}\frac{|V|}{\sigma_1 \sigma_2}\right) \exp \left( \frac{2}{\left(1-|\rho|^2\right)}\frac{\mathrm{Re}\left[V \rho^\ast \right]}{\sigma_1 \sigma_2}\right)\\
\nonumber &\equiv  \frac{1}{2\pi} \frac{\left(1 - |\rho^2| \right)}{a^2} \text{K}_0\left( \frac{|V|}{a} \right) \exp \left( \frac{\mathrm{Re}\left[V \rho^\ast \right]}{a} \right).
\end{align}
Here, we have introduced the scale parameter $a \equiv  \frac{\left(1-|\rho|^2\right)}{2} \sigma_1 \sigma_2$ for convenience. The PDF is written with respect to the standard complex metric $d\mathrm{Re}[V] d\mathrm{Im}[V]$.

The characteristic function of this visibility PDF is then given by
\begin{align}
\label{eq::char_V}
\varphi(k_r,k_i;a,\rho) = \frac{1-|\rho|^2}{1 + (a k_r - i \mathrm{Re}[\rho])^2 + (a k_i - i \mathrm{Im}[\rho])^2 },
\end{align}
where $k_r$ and $k_i$ are conjugate variables to $\mathrm{Re}[V]$ and $\mathrm{Im}[V]$, respectively.

We can calculate the PDF of the average of $N$ i.i.d.\ visibilities by inverting the product of their characteristic functions:
\begin{align}
\label{eq::V_Niid}
P(V;a,\rho,N) = \frac{1}{2^N \pi (N-1)!} \left( \frac{N}{a} \right)^{N+1}  \left( 1 - |\rho|^2 \right)^N |V|^{N-1} \text{K}_{N-1}\left( N \frac{|V|}{a} \right) \exp \left( N \frac{\mathrm{Re}\left[V \rho^\ast \right]}{a} \right).
\end{align}

If the averaged visibilities are not statistically isotropic, then we can still obtain a useful reduction using Feynman parameters to symmetrize the product of characteristic functions. These parameters $\{ s_i \}$ are defined and applied as \citep{Sred_QFT}:
\begin{align}
\frac{1}{A_1\ldots A_N}=\int dF_N(s_1A_1 + \ldots + s_N A_N)^{-N},\\
\nonumber \int dF_N=(N-1)!\int_0^1 ds_1 \ldots ds_N \, \delta(s_1+\ldots + s_N - 1).
\end{align}
The result of Eq.\ \ref{eq::V_Niid} can then be applied to the symmetrized integrand.

Because the visibility is complex, the convolution of $N$ visibilities requires a $2(N-1)$-dimensional integral. However, to symmetrize the convolution requires a single Feynman parameter for each visibility. The Feynman parameters also have an overall $\delta$-function constraint, so the convolution is reduced to an $(N-1)$-dimensional integral. If some pairs of visibilities are i.i.d., then the dimensionality of the integral can be further reduced. \citet{Vela_18cm} applied this reduction, in conjunction with the Gaussian approximation (\S\ref{sec::Gaussian_Approximation_V}), to evaluate the effects of an extended emission region on the PDF of visibility; they thereby reduced the dimensionality of the necessary numerical integration from four dimensions to one.

\subsection{Visibility Projections}
\label{sec::Visibility_Projections}
We now calculate the real and imaginary projections of the visibility PDF (Eq.\ \ref{eq::V_Niid}). We again utilize the characteristic function: the conjugate projected variable is set to zero, and the remaining function is inverted with respect to the unprojected variable. If the averaged visibilities are i.i.d., we obtain
\begin{align}
\label{eq::P_proj}
P(V_{\rm r};a,\rho,N) = \frac{N}{(N-1)!}  \frac{\sqrt{1-\rho_{\rm i}^2}}{a}   \left[ \frac{N}{2}\left(\frac{1-|\rho|^2 }{\sqrt{1 - \rho_{\rm i}^2}}\right) \frac{|V_{\rm r}|}{a} \right]^{N} k_{N-1}\left( N \sqrt{1-\rho_{\rm i}^2} \frac{|V_{\rm r}|}{a} \right) e^{N \rho_{\rm r} \frac{V_{\rm r}}{a}}.
\end{align}
Here, $\rho_{\rm r} \equiv \mathrm{Re}(\rho)$, $\rho_{\rm i} \equiv \mathrm{Im}(\rho)$, $V_{\rm r} \equiv \mathrm{Re}(V)$, and $k_N(x)$ is the modified spherical Bessel function of the second kind \citep{Arfken}. The imaginary projection $P(V_{\rm i};a,\rho,N)$ follows from the substitutions $\rho_{\rm r} \leftrightarrow \rho_{\rm i}$ and $V_{\rm r} \rightarrow V_{\rm i}$. Setting $N = 1$ and $\rho\in \mathbb{R}$ recovers the results given in the appendix of \citet{IVSS}.

We also present the projected imaginary variance: $Q(V_{\rm r}) \equiv \int V_{\rm i}^2 P(V_{\rm r},V_{\rm i}) dV_{\rm i}$. For this calculation, we again use the characteristic function, but take two derivatives with respect to $k_i$ and multiply the result by $-1$ before zeroing $k_i$ and inverting with respect to $k_r$:
\begin{align} 
\label{eq::Q_proj}
Q(V_{\rm r};a,\rho,N) = & \frac{2}{N!} \frac{a}{\left( 1 - |\rho|^2 \right)} \left[ \frac{N}{2} \frac{\left( 1 - |\rho|^2 \right)}{\sqrt{1-\rho_{\rm i}^2}} \frac{|V_{\rm r}|}{a} \right]^{N+1}\\
\nonumber & \times \left[  N \rho_{\rm i}^2 \frac{|V_{\rm r}|}{a} k_{N+1}\left( N \sqrt{1-\rho_{\rm i}^2}\frac{|V_{\rm r}|}{a} \right) 
+ \sqrt{1-\rho_{\rm i}^2} k_{N}\left( N \sqrt{1-\rho_{\rm i}^2}\frac{|V_{\rm r}|}{a} \right) \right] e^{N \rho_{\rm r} \frac{V_{\rm r}}{a}}.
\end{align}
Additional projections can be performed similarly.

\bibliography{Visibility_Short_References.bib}

\end{document}